\documentclass[12pt]{article}
\usepackage{makeidx}
\usepackage{graphicx}
\usepackage{epsf,amsmath,bbold,amsfonts,stmaryrd}

\usepackage{appendix}
\usepackage{amssymb}
\usepackage{hyperref}
\usepackage{float}

\def\a{\alpha}
\def\b{\beta}

\def\D{\Delta}

\def\eps{\varepsilon}
\def\f{\frac}
\def\g{\gamma}

\def\G{\Gamma}

\def\l{\left}

\def\mc{\mathcal}

\def\m{\mu}

\def\nn{\nonumber}

\def\p{\partial}

\def\r{\right}

\def\t{\theta}

\def\Tr{\mathrm{Tr}}

\def\x{\xi}

\def\z{\zeta}

\def\be{\begin{equation}}
\def\ee{\end{equation}}

\def\bea{\begin{eqnarray}}
\def\eea{\end{eqnarray}}

\def\ba{\begin{array}}
\def\ea{\end{array}}

\def\bc{\begin{center}}
\def\ec{\end{center}}

\def\bl{\begin{flushleft}}
\def\el{\end{flushleft}}

\def\br{\begin{flushright}}
\def\er{\end{flushright}}

\def\bi{\begin{itemize}}
\def\ei{\end{itemize}}

\def\bt{\begin{tabular}}
\def\et{\end{tabular}}

\begin{document}

\begin{titlepage}
\vspace{5cm}

\vspace{2cm}

\begin{center}
\bf \Large{Partition function on spheres: how (not) to use zeta function regularization}

\end{center}

\begin{center}
{\textsc {Alexander Monin}}
\end{center}

\begin{center}
{\it Institute of Physics, \\
\'Ecole Polytechnique F\'ed\'erale de Lausanne, \\ 
CH-1015, Lausanne, Switzerland}
\end{center}

\begin{center}
\texttt{\small alexander.monin@epfl.ch} 
\end{center}

\vspace{2cm}

\begin{abstract}

It is known that not all summation methods are linear and stable. Zeta function regularization is in general non-linear. However, in some cases  formal manipulations with "zeta function" regularization (assuming linearity of sums) lead to correct results. We consider several examples and show why this happens.

\end{abstract}

\end{titlepage}

\section{Intro and motivation}

It should be stated right away that {\it a priori} there is no right or wrong regularization scheme. No physical observable should depend on the choice of the procedure, as long as it qualifies as a regularization, meaning that there exists a smooth limit of zero regulator resulting in a local theory with a proper particle/field content and symmetries. 

One regularization can be more appropriate than the other in the sense that it allows to reproduce the necessary (specific) Ward identities with lesser effort. In particular symmetry preserving regularizations are usually favored among others, however, one is not forced by any means to make this choice, and if done carefully, computations using different regularizations, lead to the same predictions for physical observables. 

One of the regularizations suitable for computing determinants of differential operators (one-loop partition functions) or Casimir energies is zeta function regularization. Even though the subtleties of its uses have been discussed on numerous occasions (for systematic approach see for example~\cite{Elizalde:1994gf, Vassilevich:2003xt, Dunne:2007rt, Elizalde:2012zza}) there are still instances when it is handled in a somewhat cavalier way leading to inconsistencies. We do not claim that results obtained in these cases are wrong, it is rather the matter of interpreting those results. 

If a regularization breaking symmetries of a system is used, one may be forced to introduce non-symmetric counterterms to compensate for it in the Ward identities. It means that the space of possible operators in the Lagrangian is not restricted by those respecting the symmetries.
The feature of zeta function regularization is that it is diffeomorphism invariant. Thus, the analysis boils down to listing all possible diffeomorphism invariant operators in the Lagrangian. However, if the regularization is not used in a proper way there is no guarantee that the diffeomorphisms play any role in restricting the terms in the Lagrangian. Thus, the analysis with only diffeomorphism invariant terms may be not complete, leading to a non-complete result.

In this paper we consider a scalar field coupled to a round sphere. We show several examples of how to compute its free energy using different regularizations, like zeta function, Pauli-Villars, the heat kernel, and show that all of them produce the same result. Surprisingly enough sometimes a misuse of zeta function regularization leads to a correct prediction for a finite part of the free energy. We show examples when that happens and explain the reason. Some of our results are necessarily new and can be found in cited textbooks, but we hope 
that our presentation and new results provide certain clarification.

The paper is organized as follows. In Section~\ref{zeta_function_intro} we define different zeta function to be used later on. In Sections~\ref{section_PV} and~\ref{section_zeta-function} we show how to use Pauli-Villars and zeta function regularizations correspondingly to compute the free energy of a scalar on a round sphere. In Section~\ref{section_t-expansion} we present yet another regularization using a smoothed cut-off function. Section~\ref{section_examples} is dedicated to examples and in Section~\ref{section_conclusion} we present our conclusions. Some relevant formulas and more detailed computations can be found in Appendices.

\section{Different zeta functions \label{zeta_function_intro}}

In this section we introduce the definitions of different zeta functions along with some of their properties which will be useful in sections to come.

\subsection{Riemann zeta function}

Riemann zeta function $\z (s)$ is defined for $\mathrm {Re} \, s >1$ as an infinite convergent sum
\be
\z (s)=\sum _ {k=1} ^ \infty \f {1} {k^s}.
\label{Riemann_zeta}
\ee
For $s<1$ one defines zeta function using analytic continuation. From the integral representation of the $\G$-function
\be
\G (s) = \int _ 0 ^ \infty dt e ^ {-t} t ^ {s-1}, ~~ \mathrm {Re} \, s >1,
\ee
it is easy to show that for $\mathrm {Re} \, s >1$ the following relation holds
\be
\zeta (s) \G (s) = \int _ 0 ^ \infty dt \, \t (t) \, t ^ {s-1},
\label{zeta_gamma_continuation}
\ee
where we introduced the function $\t(t)$ (later to be called the heat kernel) as the sum  
\be
\t (t) = \sum _ {k=1} ^ \infty e ^ {-k t} = \f {1} {e ^ t -1}  
\ee
convergent for all positive values of $t$. Breaking up the integral in (\ref{zeta_gamma_continuation}) into two parts and doing a trivial integration
\be
\int _ 0 ^ 1 + \int _ 1 ^ \infty \f {t ^ {s-1} \, d t} {e ^ t -1} = 
\int _ 0 ^ 1 t ^ {s-1} \, d t \l ( \f {1} {e ^ t -1} - \f {1} {t} \r ) + \f {1} {s - 1} + \int _ 1 ^ \infty \f {t ^ {s-1} \, d t} {e ^ t -1},
\ee
one gets the expression for the zeta function which can be used for $\mathrm{Re} s > 0$
\be
\z (s) \G (s) -  \f {1} {s - 1} = \int _ 0 ^ 1 t ^ {s-1} \, d t \l ( \f {1} {e ^ t -1} - \f {1} {t} \r ) + \int _ 1 ^ \infty \f {t ^ {s-1} \, d t} {e ^ t -1}, ~~ 
\mathrm{Re} (s) > 0.
\ee
The function can be analytically continued to the region $\mathrm{Re} \, s > - N$, where $N$ is an arbitrary positive integer, in a similar manner.

\subsection{Hurwitz zeta function}

A natural generalization of (\ref{Riemann_zeta}) is the Hurwitz zeta function defined as
\be
\z (s,q)=\sum _ {k=0} ^ \infty \f {1} {(k+q)^s}, ~~ \mathrm {Re} \, s >1.
\label{Hurwitz_zeta}
\ee
Analytic continuation of the Hurwitz zeta function is obtained analogously to the one described in the previous section. Using the Mellin transform
\be
\Phi (z) = \int _ 0 ^ \infty x^ {z-1} (x+q) ^{-s} dx = q ^ {z-s} \f {\G (s-z) \G (z)} {\G (s)},
\ee
which is holomorphic in the strip $ 0 < \mathrm {Re} \, z <  \mathrm {Re} \, s $ one can rewrite for $c > 1$ and $\mathrm {Re} \, s > 1$ the zeta function in the following form
\be
\z (s,q) = q^{-s} + \sum _ {k=1} ^ \infty \f {1} {(k+q)^s} = \f {1} {2 \pi i} \int \limits _ {c - i \infty} ^ {c + i \infty} \sum _ {k=1} ^ \infty k ^ {-z}
\Phi(z) dz =  \f {1} {2 \pi i} \int \limits _ {c - i \infty} ^ {c + i \infty}  \z (z) \Phi(z) dz.
\ee
Closing the contour of integration in the left half plane the integral can be written as a sum over residues
\bea
\label{asymptotic_Hurwitz_zeta}
\z (s,q) & = & \int _ 0 ^ \infty (x+q) ^{-s} dx + 
\f {q ^ {-s}} {\G (s)} \sum _ {k=0} ^ \infty \mathrm {res} \l [ q^ z \G (s-z) \G (z) \z(z), z= - k \r ] \\
& = & - \f {q ^ {1-s}} {1-s} + \f {q ^ {-s}} {2} + \sum _ {m=1} ^ \infty \f {B _ {2m}} {(2m)!} q ^ {1-s - 2m} s (s+1) \dots (s + 2 m - 2), \nn
\eea
where the first integral in (\ref{asymptotic_Hurwitz_zeta}) corresponds to the pole of $\z(s)$ at $s=1$. This expansion for the zeta function should be understood as an asymptotic one. The result can be used to find a large $L$ expansion of the following finite sum
\be
\sum _ {n =1} ^ L \f {1} {(n+q)^s} = \z (s,q + 1) - \z (s,L+q+1), ~~ s \neq 1,
\label{finite_two_zeta}
\ee
which will be used in the following sections.

\subsection{Spectral zeta functions \label{heat_kernel_section}}

In general for a differential operator $D$ with positive eigenvalues $\Lambda _ \a > 0$ one defines spectral zeta function as the sum
\be
\z _ D (s) = \sum _ \a \Lambda _ \a ^ {-s} \equiv \mathrm {Tr} D ^ {-s},
\label{zeta_spectral}
\ee
for sufficiently large $\mathrm {Re} \, s > s _ 0$. The analytic continuation is performed using the corresponding heat kernel, which is given by
\be
\t _ D (t) = \sum _ \a e ^ {- t \Lambda _ \a} \equiv \Tr e ^ {-D t}.
\label{heat_kernel}
\ee
If the asymptotic expansion of $\t (t)$ for $t \to 0$ is known
\be
\t _ D (t) \underset{t \to 0} {=} \sum _ {k \geq 0} a _ k t ^ {-s_ k} + O (t),
\label{heat_exp_gen}
\ee
with $s_0>s_1>s_2>\dots>0$ and $a_k$ being constants, then the analytic continuation is done in exactly the same way as for the Riemann zeta function. For instance, the following expression
\be
\z _ D (s) \Gamma (s) =  \f {a _ 0} {s-s_0} + \int _ 0 ^ 1 dt \, \l ( \t _ D (t) - \f{a_0}{t ^ {s_ 0}} \r ) t ^ {s-1} + \int _ 1 ^ \infty dt \, \t _ D (t) t ^ {s-1},
\label{zeta_analytic_cont}
\ee
can be used to define the spectral zeta function of the differential operator $D$ in the domain~$\mathrm {Re} \, s > s _ 1$. It is clear that the powers $s _ k$ in (\ref{heat_exp_gen}) are in one to one correspondence with the poles of the zeta function $\z_D(s)$. In general after analytic continuation the zeta function can be written as
\be
\z _ D (s) \G (s) = G _ D (s),
\ee
where $G _ D (s)$ is the analytic continuation of the r.h.s. of (\ref{zeta_analytic_cont}). It is straightforward to show that around $s=0$ the following expansion -- provided $\z _ D (0)$ is regular -- holds
\be
G _ D (s) \underset{s\to 0}{=} \f {\z _ D (0)} {s} + \z ' _ D (0) - \g \z _ D (0) + \dots,
\label{G_s_expansion}
\ee
hence $\z _ D (0)$ can be found as a $t$-independent (compare with (\ref{gen_theta_asympt})) coefficient in (\ref{heat_exp_gen}).

\section{Pauli-Villars regularization \label{section_PV}}

Let us consider a free scalar field on a $d$-dimensional Euclidean sphere (not necessarily minimally coupled to a round metric) and compute its partition function. Computations of the free energy of a scalar on spheres can be found in several papers, for instance the case of conformally coupled scalar was considered in~\cite{Klebanov:2011gs}. Since the Ricci curvature is constant for a round sphere $R = d(d-1)/a^2$, the action for both massive and massless scalar on a sphere can be written in a general form as
\be
S[\phi,g] = \f {1} {2} \int d ^ d x \sqrt{g} \l [ \l ( \nabla \phi \r ) ^ 2 + \a _ m ^2 (\x) \phi ^ 2 \r ],
\label{bare_action}
\ee
where $\a _ m ^2 (\x)= m^2 + \x R$ with $\xi$ being a constant and $a$ the radius of the sphere. For a specific value ($d > 1$)
\be
\x = \f {d-2}  {4(d-1)}
\label{xi_conformal}
\ee
the action becomes Weyl invariant (conformally coupled scalar). The free energy $F$ is by definition a logarithm of the partition function
\be
Z = e ^ {-F} = \int \mc D \phi e ^ {-S[\phi,g]}, ~~ \text{or} ~~ F = \f {1} {2} \log \det \Delta,
\label{formal_det}
\ee
where the operator $\D$ is given by
\be
\D = - \nabla ^ 2 + \a _ m ^ 2(\x).
\label{general_delta}
\ee
For an $n$-dimensional sphere this operator has eigenvalues
\be
\lambda _ \ell = \f {\ell (\ell+d-1)} {a^2} + \f {\a _ m ^2(\x)} {a^2},
\ee
with multiplicity
\be
\mu _ \ell = \f {(2\ell + d-1)(\ell+d-2)!} {(d-1)!d!}, ~~ \ell = 0,1,\dots
\label{degeneracy}
\ee
It is clear that the formal expression (\ref{formal_det}) is divergent and needs regularization. Before showing how zeta functions can be used to compute the free energy we make use of an alternative regularization. In this way we will have a reference point for the computations involving zeta functions. One of the most straightforward regularizations is the Pauli-Villars (PV) one, in which the action (\ref{bare_action}) is regulated by adding sufficient number of regulator fields with mass $M_i$ and statistics $-c_i$ (in general fractional). It can also be viewed as adding higher derivative operators.

\subsection{$S^1$ \label{PV_S1_ex}}

For a one dimensional sphere to regularize the action
\be
S = \f {1} {2} \int d t \l ( \dot \phi ^ 2 + m^ 2 \phi ^ 2 \r ),
\ee
it is enough to introduce one PV regulator with mass $M$ and opposite to $\phi$ statistics. 
The corresponding eigenvalues are
\be
\lambda _ \ell = \f {\ell^2} {a^2}+m^2, ~~ \text{and} ~~ \lambda ^ M _ \ell = \f {\ell^2} {a^2} + M ^ 2, ~~ \ell \in \mathbb Z.
\label{lambda_S1}
\ee
As a result the expression for the regularized free energy reads
\be
F ^ R _ {S^2} = \f {1} {2} \log \prod _ {\ell \in \mathbb Z} \f {\lambda _ \ell } {\lambda _ \ell ^ M}
= \log \l ( \f {m} {M} \prod _ {\ell =1} ^ \infty \f { \ell ^ 2 + m^2 a ^ 2} {\ell ^ 2 + M ^ 2 a^ 2} \r ) = \log \l ( \f{ \sinh {\pi a m} } {\sinh {\pi a M}} \r ),
\ee
which upon the limit $M \to \infty$ becomes
\be
F ^ R _ {S^2} = \log \l ( 2 \sinh {\pi a m}  \r ) - \pi a M.
\label{PV_S2}
\ee
The divergent part can be eliminated by the counterterm $\mathrm {Vol} _ {S ^ 2} = 2 \pi a$ with an appropriate coefficient.

\subsection{$S^3$}

In the case of a three dimensional sphere the free energy can be regularized employing only two PV fields
\be
2 F ^ R _ {S^3} = \sum _ {\ell = 1} ^ \infty \ell ^ 2 \l \{ \log \l [ \ell ^ 2 - 1 + \a _ m ^ 2(\x) a^2 \r ] - 
\sum _ {i=1} ^ 2 c _ i \log \l [ \ell ^ 2 - 1 + \a _ {m} ^ 2(\x) a^2 + M^2_i a^2 \r ] \r \},
\label{FS3}
\ee
with mass parameters satisfying the following relations
\bea
c_1 + c _ 2 & = & 1 \nn \\
c _ 1 M _ 1 ^ 2 + c _ 2 M _ 2 ^ 2 & = & 0.
\label{ci_constr}
\eea
Introducing the cutoff $L \gg M _ i a \gg 1 $ the sum with PV regulators can be computed using Euler-Maclaurin summation formula, while 
the first term in (\ref{FS3}) can be found using the zeta function expansion (\ref{asymptotic_Hurwitz_zeta}) and the formula (\ref{finite_two_zeta}). As a result for $M _ 1 = M$ and $M _ 2 = r M$ with $r$ being a number we obtain the following expression for a regularized partition function in the limit $M \to \infty$
\be
F ^ R _ {S^3} = - a^3 M ^ 3 \f {\pi r ^ 2} { 6 (1+r) }  - a M \f {\pi r} {4( 1 + r )} \l [1-\a_m^2(\x) a^2 \r ] + F _ {S ^ 3} \l (\sqrt{\a _ m ^2(\x) a ^ 2-1}\r ).
\label{PV_S3}
\ee
Divergent terms can be removed by the counterterms of the form
\be
b_1 M ^ 3 \mathrm {Vol} _ {S ^ 3} + b_2 M \mathrm {Ric} _ {S ^ 3},
\label{counterterms _S3}
\ee
with appropriate coefficients $b_{1,2}$, where $ \mathrm {Ric} _ {S ^ 3}$ is the Ricci curvature. The function $F_ {S^3}(x)$ is given in the Appendix~\ref{Appendix_PV}. For a limiting case of conformal coupling and zero mass it becomes
\be
f _ {S ^ 3} = \f {1} {16} \l ( 2 \log 2 - \f {3 \z (3)} {\pi ^ 2} \r ) = 0.0638,
\label{f_conf_scalar}
\ee
which coincides with the result for the partition function from~\cite{Klebanov:2011gs}.

\section{Zeta function regularization \label{section_zeta-function}}

As was discussed in the introduction different regularizations should lead to one and the same prediction for an observable. However, different procedures (schemes) allow different number of counterterms. For instance, in the previous section we considered the PV regularization which obviously preserves the diffeomorphisms but even in the case of zero mass $m=0$ and conformal coupling (\ref{xi_conformal}) it breaks Weyl invariance due to the presence of a scale parameter (regulator mass). This is precisely what we see from the expressions for the regularized partition function (\ref{PV_S2}) and (\ref{PV_S3}), where the divergent terms are clearly of a diffeomorphism invariant nature but are not Weyl invariant.

In general for the case of zero mass $m=0$ using a regularization preserving the diffeomorphisms but having a scale $M$ the only local counterterms one can write are of the form
\be
M ^ d \mathrm {Vol}, ~ M ^ {d-2} \mathrm {Ric}, ~ M ^ {d-4} \mathrm {Ric^2}, \dots
\ee
Thus the power counting tells us that in an odd\footnote{In an even number of dimensions $\log$ divergent terms lead to the Weyl anomaly.} number of dimensions $d$ a counterterm scaling as $M^0$ does not exist, hence, the $M$-independent (finite) part of the regularized partition function is a prediction and can be associated with the (renormalized) partition function of a scalar on a $d$-sphere $f _ {S ^d}$. There is simply no counterterm that may change it. It is not the case for a massive scalar, for even $M$-independent part can be changed using counterterms, for example $m ^ d \mathrm {Vol}$.

If instead for a conformally coupled scalar (still odd number of dimensions) a regularization preserving the Weyl invariance was used, the result for the partition function would be produced automatically without any need to add counterterms for there are none satisfying the constraints of Weyl invariance. To show that we consider a bit more general setup. In this way we will also obtain the expression for the partition function in terms of the zeta function of the corresponding operator. To find a partition function of a scalar field living on a $d$-dimensional manifold $\mathbb M^d$, instead of introducing PV regulators we employ dimensional regularization and put a scalar on
$\mathbb M ^ d \times \mathbb R ^ {2 \eps}$, in the end the limit $\eps \to 0$ should be taken. It follows from (\ref{formal_det}) that the regularized partition function has the form
\be
F ^ R = -\f {1} {2} \f {\p} {\p s} \Big | _ {s=0} \sum _ {\a} \int \f {d^ {2\eps}x \, d^ {2\eps} k} {(2 \pi) ^ {2 \eps}} 
\l [ \f {k ^ 2} {\m ^ 2} + \f {\Lambda _ \a} {\m ^ 2} \r ] ^ {-s},
\ee
where $\Lambda _ \a$ are eigenvalues of the operator of second variation and $\m$ is a normalization scale. Computing the integral we obtain
\be
F ^ R = -\f {1} {2} \f {\p} {\p s} \Big | _ {s=0} \f {\G(s-\eps)} {\G(s)} \sum _ {\a} \l ( \f {\mu ^ 2 L ^ 2} {4 \pi} \r ) ^ \eps
\l ( \f {\Lambda _ \a} {\m ^ 2} \r ) ^ {\eps-s},
\ee
where $L$ is the size of an auxiliary space (infrared regulator). The above expression should be understood in the sense of an analytic continuation in $\eps$. Equivalently one can view it as an analytic continuation in $s$ and set $\eps=0$ (there are no $1 / \eps$ poles in odd number of dimensions). As a result we get
\be
F ^ R = -\f {1} {2} \f {\p} {\p s} \Big | _ {s=0} \sum _ \a \l ( \f{\Lambda _ \a} {\m ^ 2} \r ) ^ {-s} = - \f {1} {2} \z ' _ D (0) 
- \f {1} {2} \z _ D (0) \log \m ^ 2,
\label{F_zeta_function}
\ee
where $\z _ D (s)$ is nothing else but the corresponding zeta function (\ref{zeta_spectral}). 

In order to compute a partition function of a conformally coupled scalar on a round $d$-dimensional sphere
we first have to find a small $t$ asymptotic expansion of the corresponding heat kernel
\be
\t_ {S ^ d} (s) = \sum _ {\ell=0} ^ \infty \m _ \ell \exp \l [ - t \l(\ell + \f {d} {2} \r) \l(\ell + \f {d} {2} -1 \r) \r ].
\ee
Using the Euler-Maclaurin formula (see Appendix \ref{Appendix_PV}) it is easy to show that
\be
\t_ {S ^ d} (s) \underset{t\to 0}{=} \sum _ {k=0} ^ {(d+1)/2}a _ {d/2-k} t ^ {-d/2+k}+ O (t).
\ee
Hence we get
\bea
\z _ {S ^ d} (s) \Gamma (s) = G _ {S ^ d}(s) \equiv & \displaystyle \int  _ 0 ^ \infty & dt \, \l [ \t _ {S ^ d} (t) 
- \sum _ {k=0} ^ {(d+1)/2}a _ {d/2-k} t ^ {-d/2+k} \r ] t ^ {s-1} \\
& + &\sum _ {k=0} ^ {(d+1)/2} \f { a _ {d/2-k} } {s - d/2 + k} + \int _ 1 ^ \infty dt \, \t _ {S ^ d} (t) t ^ {s-1}. \nn
\eea
Since there is no pole at $s=0$ it means that for an arbitrary odd $d$ the zeta function vanishes $\z _ {S ^ d} (0)=0$. In this case the partition function of a conformally coupled scalar is given by
\be
f _ {S ^ d} = - \f {1} {2} \f {\p} {\p s} \Big | _ {s=0} \sum _ {\ell=0} ^ \infty \m _ \ell \l [ \l(\ell + \f {d} {2} \r) \l(\ell + \f {d} {2} - 1 \r) \r ] ^ {-s} (\m^2 a^2) ^ s
= - \f {1} {2} \z ' _ {S ^ d} (s),
\ee
and it does not depend on the normalization scale $\m$. Just as an example for $3$-dimensional sphere we get
\be
G _ {S ^ 3}(0) = \int _ 0 ^ \infty \f{dt} {t} \, \l [ \t _ {S ^ 3} (t) - \sqrt{\pi } \l ( \frac{1}{4 \, t^{3/2}}+\frac{1}{16 \, t ^ {1/2}} + \frac{t ^ {1/2}}{128} \r )  \r ] - \f {53\sqrt{\pi }} {192} + \int _ 1 ^ \infty \f{dt} {t} \, \t _ {S ^ d} (t).
\label{S3_zeta_function_reg}
\ee
Computing the integrals numerically we obtain the same value as in (\ref{f_conf_scalar}).

\subsection{How not to use zeta function regularization}

We should stress that although it is tempting to continue the formal manipulations with (\ref{formal_det}), it is the expression (\ref{F_zeta_function}) that we obtained for the partition function using the dimensional (in the case at hand it is reduced to the zeta function) regularization and not something else. It is easy to face an inconsistency if the regularization is not used with proper care. We will illustrate it on the example of a $3$-sphere with a conformally coupled massless scalar. The formal expression for the partition function is given by\footnote{We deliberately use capital letter $F _ {S ^ 3}$ to distinguish from the correct expression given by $f _ {S ^ 3}$.}
\bea
F_ {S^3} & = &
\f {1} {2} \sum _ {\ell=0} ^ \infty (\ell + 1)^2 \l \{ \log \l [ \l (\ell + \f {3} {2} \r)\l( \m a \r )^p \r ] + \log \l [ \l(\ell + \f {1} {2} \r)\l( \m a \r ) ^ {2-p} \r ] \r \} \nn \\ 
& = & - \f {1} {2} \f {\p} {\p s} \Big | _ {s=0} \l [ (\m a) ^ {-ps} \sum _ {\ell=0} ^ \infty (\ell + 1 ) ^ 2 \l (\ell + \f {3} {2} \r)^{-s} + (\m a) ^ {-(2-p)s} \sum _ {\ell=0} ^ \infty (\ell + 1)^2 \l (\ell + \f {1} {2} \r)^{-s} \r ]  \nn \\
 & = & - \f {1-p} {24}\log \l( \m a \r ) + \f {1} {16} \l ( 2 \log 2 - \f {3 \z (3)} {\pi ^ 2} \r ),
 \label{f_formal}
\eea
where $p$ is an arbitrary constant. The term $\log (\m a)$ signals the presence of a $\log$-divergent term, meaning that -- since there are no $\log$-divergent diffeomorphism invariant counterterms -- the regularization breaks diffeomorphisms. Even for $p=1$ when (\ref{f_formal}) reproduces (\ref{f_conf_scalar}) these manipulations are illegitimate because
\be
\sum _ {\ell=0} ^ \infty \f{(\ell + 1)^2}  {\l (\ell + \f {3} {2} \r)^{s}} + \sum _ {\ell=0} ^ \infty \f{(\ell + 1)^2}  {\l (\ell + \f {1} {2} \r)^{s}} \neq
\sum _ {\ell=0} ^ \infty \f{(\ell + 1 ) ^ 2} {\l [ \l(\ell + \f {3} {2} \r) \l(\ell + \f {1} {2} \r) \r ] ^ {s}}, ~~ s > 3.
\ee
So it is a wrong function whose analytic continuation to $s=0$ we use in (\ref{f_formal}). 

From a purely mathematical perspective it can be understood in the following way. There are different notions of summation like Abel, Ces\'aro, Borel etc. (see~\cite{Hardy}) that are used to give meaning to otherwise divergent series. However, it happens so that some divergent series, for instance
\be
1+2+3+\dots,
\ee
cannot be summed using linear and stable methods\footnote{Linear means that the sum of two series is equal to the sum of the series obtained as a combination of the two, and stable means that adding a number to the series increases the sum by the same amount.}. The series in question $\sum _ \ell \m _ \ell \log \lambda _ \ell$ is one of them, thus the formal step $\log A B = \log A + \log B$ in the sum is inconsistent.

Another example -- considered in~\cite{Brevik:1989fw,Elizalde:1996zk} -- when formal manipulations lead to a wrong result is the computation of the Casimir energy of a piecewise uniform, closed string. The formal expression for the energy is given by
\be
E ^ \text{formal} _ \text{Cas} = \sum _ {k=0} ^ \infty \eps _ k = \sum _ {k=0} ^ \infty \l ( k + \b \r ).
\label{Casimir_string}
\ee
If regularized with the zeta function the energy can be expressed in terms of the Hurwitz zeta function
\be
E _ \text{Cas} = \lim _ {s \to -1} \sum _ {k=0} ^ \infty \eps _ k ^ {-s} = \z (-1,\b),
\ee
which is clearly different from the formal substitution
\be
E ^ \text{formal} _ \text{Cas} = \sum _ {k=0} ^ \infty \l ( k + \b \r ) = \z (-1) + \b \l [ 1+ \z (0) \r ].
\label{Casimir_string_wrong}
\ee
At the same time imagine that the following sum has to be computed
\be
\sum _ {\ell = 1} ^ \infty \lambda _ \ell a ^ 2,
\label{zeta_1_S1}
\ee
with $\lambda _ \ell$ from (\ref{lambda_S1}). Then both methods produce the same result. Indeed, on the one hand formally we have
\be
\sum _ {\ell = 1} ^ \infty \l ( \ell ^ 2 + m ^ 2 a ^ 2 \r ) = \z (-2)+ \z(0) m^ 2 a ^ 2 = -\f {1} {2} m^ 2 a ^ 2,
\label{Epstein_S1}
\ee
on the other hand using the expression from~\cite{Elizalde:1996zk} for the Epstein zeta function $\z _ E (s,q)$ 
\be
\lim _ {s \to - 1} \sum _ {\ell = 1} ^ \infty \l ( \ell ^ 2 + m ^ 2 a ^ 2 \r ) ^ {-s} = \z _ E (-1,m^2a^2) = -\f {1} {2} m^ 2 a ^ 2,
\label{naive_S1}
\ee
leads to the same result.

\section{Heat kernel and zeta function \label{section_t-expansion}}

The zeta function regularized expression (\ref{F_zeta_function}) for the free energy (the determinant of a corresponding operator) can be simply postulated. In the previous section we showed how to justify this definition using the dimensional regularization.
However, the dimensional regularization also may involve rather formal manipulations with integrals that do not converge for any real number of dimensions\footnote{Usually the problem is solved by splitting the integral into IR and UV pieces, computing them separately and analytically continuing the result.}. Therefore, in this section we will present yet another derivation using a more physically motivated cut-off regularization. Keeping in mind that we want to preserve the diffeomorphisms it is not a good idea to use the hard cut-off. Instead we employ a smoothed cut-off for regularization~(see \cite{Dietz:1982uc})
\be
\mathrm {Reg} \, \sum _ \a h(\Lambda _ \a) \equiv \sum _ \a h(\Lambda _ \a) \eta (t \Lambda _ \a),
\label{eta_reg}
\ee
where $h(x)$ is an arbitrary function and $\eta(x)$ is a cutoff function, such that $\eta (0) = 1$ and it decays sufficiently fast at infinity
\be
\eta(x) \underset{x\to \infty} \to 0,
\ee
so that the sum (\ref{eta_reg}) is convergent. We will be interested in a very specific class of functions $h(x)$, namely $h(x) = x ^ {-s}$ or $h(x) = (x + q) ^ {-s}$ with $q$ being a constant. The claim is that for the asymptotic $t \to 0$ expansion of the sum the $t$-independent piece does not depend (see Appendix~\ref{Riemann_Hurwitz_heat} for examples) on the function $\eta(x)$ and is given by the corresponding zeta function $\z _ D(s)$. It is obvious for sufficiently large $s$ when the sum $\sum _ \a \Lambda ^ {-s} _ \a$ is convergent. We will show that for a specific choice
\be
\eta (x) = e ^ {-x},
\ee
which is convenient for practical reasons. In this case defining
\be
\t _ D (s,t) \equiv \sum _ \a \Lambda _ \a ^ {-s} e ^ {- t \Lambda _ \a},
\ee
which for $h(x)=1$ coincides with the heat kernel defined previously (\ref{heat_kernel}), one can show similarly to (\ref{heat_kernel_section}) that for sufficiently large $s'$ the following relation holds
\be
\z _ D(s+s') \G (s') = \int _ 0 ^ \infty dt \, \t _ D (s,t) t ^ {s ' - 1}.
\label{zeta_heat_rel_gen}
\ee
For small $t$ the function $\t _ D (s,t)$ can be expanded in the asymptotic series\be
\t _ D (s,t) \underset{t \to 0} {=} \sum _ {k \geq 0} a _ k (s) t ^ {-s_ k(s)} + O (t).
\label{generalized_heat}
\ee
Then the analytic continuation of (\ref{zeta_heat_rel_gen}) to a region including $s'=0$ is given by
\bea
\z _ D(s+s') \G (s') = & \displaystyle \int _ 0 ^ 1 & dt \, \l [ \t _ D (s,t) - \sum _ {s _ k \geq 0} a _ k (s) t ^ {-s_ k(s)} \r ] t ^ {s ' - 1} \\
& + &
\f {a _ {k_ 0}} {s'} + \sum _ {s _ k > 0} \f {a _ k (s)} {s' + s _ {k} (s)} +  \int _ 1 ^ \infty dt \, \t _ D (s,t) t ^ {s ' - 1}.
\eea
On the l.h.s the residue of the pole at $s'=0$ equals $\z _ D (s)$, while on the r.h.s. it corresponds to the coefficient $a _ {k_0}$ for which 
$s_{k_0}(s)=0$. Hence, we proved that
\be
\t _ D (s,t) \underset{t \to 0}{=} \sum _ {s_k(s)>0} a _ k (s) t ^ {-s _ k (s)} + \z _ D (s) + \sum _ {\tilde s_k(s)>0} \tilde a _ k (s) t ^ {s _ k (s)}.
\label{gen_theta_asympt}
\ee
This expression gives us yet another way of computing the zeta function of a differential operator. At the same time it gives a more physical perspective on the formula (\ref{F_zeta_function}). All terms except for the $t$-independent one depend on the choice of the cut-off function 
$\eta (x)$ (see the Appendix~\ref{Riemann_Hurwitz_heat}). However, the terms in (\ref{gen_theta_asympt}) with positive powers of $t$ are irrelevant for the limit $t \to 0$ and those with negative powers of $t$ can be removed with a proper choice of local counterterms, leaving the $t$-independent term (if there are no counterterms scaling as $t^0 \sim a^0$, i.e. there is no $\log$ in (\ref{F_zeta_function})) as a prediction of an observable (for instance free energy or Casimir energy).

\section{Examples \label{section_examples}}

\subsection{$S ^ 1$}

The regularization discussed in the previous section makes it clear why sometimes obviously not legitimate operations -- like the ones in (\ref{naive_S1}) -- lead to the same result as the analytic continuation (\ref{Epstein_S1}). Let us consider again the sum (\ref{zeta_1_S1}).
Its analogue after the regularization with a smoothed cut-off has the form
\be
\sum _ {\ell = 1} ^ \infty \l ( \ell ^ 2 + m ^ 2 a ^ 2 \r ) e ^ {- t \l ( \ell ^ 2 + m ^ 2 a ^ 2 \r )} = e ^ {- t \, m ^ 2 a ^ 2} \l ( \sum _ {\ell = 1} ^ \infty \ell ^ 2 e ^ {- t \ell ^ 2} + m ^ 2 a ^ 2 \sum _ {\ell = 1} ^ \infty e ^ {- t \ell ^ 2} \r ),
\label{ex_1}
\ee
where now one is free to regroup the sum in all possible ways, for it converges. Using the formula (\ref{Riemann_Hurwitz_heat_formula}) it is straightforward to show that the expression in braces on the r.h.s. of  (\ref{ex_1}) has the following asymptotic expansion for $t\to 0$
\be
\f{\G(3/2)} {t ^ {3/2}} + \z(-2) + m^2a^2 \l ( \f{\G(3/2)} {t ^ {1/2}} + \z(0) \r ).
\label{ex_1_rewritten}
\ee
It is clear that the factor $e ^ {- t \, m ^ 2 a ^ 2}$ cannot change the $t$-independent part in (\ref{ex_1_rewritten}) since its expansion runs only over integer powers of $t$, thus so regularized sum reproduces the result from (\ref{Epstein_S1}).

On the other hand the sum (\ref{Casimir_string}) after regularization becomes
\be
 e ^ {-t \b} \l ( \sum _ {k=0} ^ \infty k e ^ {-k t} + \b \sum _ {k=0} ^ \infty e ^ {-k t} \r )
\underset {t \to 0} {=} e ^ {-t \b} \l [ \f {1} {t^2} + \z (-1) + \b \l ( \f {1} {t} + \z (0) + 1 \r ) \r ].
\label{ex_2_rewritten}
\ee
The result (\ref{Casimir_string_wrong}) corresponds to the $t$-independent part of the expression in brackets in (\ref{ex_2_rewritten}). The term $e ^ {-t \b}$ is crucial, for it leads to the following $t$-independent part of the whole sum
\be
\z (-1) + \b \l [ 1+ \z (0) \r ] - \f {\b^2} {2} = \z (-1,\b),
\ee
coinciding with (\ref{Casimir_string}). 

Another aspect of the computation that should be stressed is the following. The regularization (\ref{ex_2_rewritten}) leads to only one divergent piece, $1/t^2$, while if one drops the term $e^{-\b t}$ -- leading to (\ref{Casimir_string_wrong}) -- there are two divergent terms, namely
\be
\f {1} {t ^ 2} + \f {\b} {t}.
\ee
It means that in order to get the finite part one has to introduce additional counterterms, which is signaling that the regularization is not covariant.

\subsection{$S ^ 3$}

Now we move to a $3$-dimensional example of a scalar field conformally coupled to a round metric. The quantity to be computed is the following
\be
f _ {S ^3} = - \f {1} {2} \z ' _ {S ^ 3} (0) = F _ {S ^ 3} (t) \Big | _ {t\text{-indep}}.
\ee
with
\be
F _ {S ^ 3} (t) = \f {1} {2} \sum _ {\ell=1} ^ \infty  \ell ^ 2 
\log \l [ \m ^ {-2} a ^ {-2} \l(\ell + \f {1} {2} \r) \l(\ell - \f {1} {2} \r) \r ] e ^ {-t \l(\ell^2 - {1} / {4} \r)}.
\ee
As we saw before (see (\ref{ex_1}) and (\ref{ex_1_rewritten})) the term with $\log \m^2 a ^2$ does not contribute to the finite $t$-independent part of the sum. Taking into account that for large $\ell$
\be
\ell^2 \log (\ell ^2 -1/4) = \ell^2 \log \ell ^ 2 - \f {1} {4} + O (\ell ^ {-2}),
\ee
we obtain
\be
f _ {S ^3} = \f {1} {2} \sum _ {\ell=1} ^ \infty \l ( \ell^2 \log \ell ^ 2 - \f {1} {4} \r ) e ^ {-t \l(\ell^2 - {1} / {4} \r)} \Big | _ {t\text{-indep}} + 
\f {1} {2} \sum _ {\ell=1} ^ \infty  \l [\ell ^ 2 \log \l(\ell^2 - \f {1} {4} \r) - \ell^2 \log \ell ^ 2 + \f {1} {4} \r ],
\label{t_expansion_zeta}
\ee
where in the second sum we dropped the factor $e ^ {- t (\ell^2-1/4)}$, for it is convergent. The last sum can be computed numerically while using the formulas~(\ref{Riemann_Hurwitz_heat_formula}) we get 
\be
\sum _ {\ell=1} ^ \infty \l ( \ell^2 \log \ell ^ 2 - \f {1} {4} \r ) e ^ {-t \ell^2} = - \f {\sqrt{\pi}} {4 \, t ^ {3/2}} \l ( \log 4 t + \g -2 \r ) - \f {\sqrt{\pi}} {8 \, t ^ {1/2}} - \f {\z (0)} {4} - 2 \z' (-2).
\label{naive_S3_log}
\ee
Hence, since the additional -- compared to (\ref{naive_S3_log}) -- factor $e ^ {t/4}$ in the first term in (\ref{t_expansion_zeta}) does not change the $t$-independent part of the sum\footnote{This is the reason why the naive computation leads to the correct result. See also the Section~\ref{section_conclusion}.}, we get the same numerical result as in~(\ref{f_conf_scalar}) and~(\ref{S3_zeta_function_reg})
\be
f _ {S ^3} = - \z ' (-2) - \f {1} {8} \z (0) + 0.02914 = 0.0638.
\ee
We find it also instructive to compute the free energy, as a $t$-independent term in the sum, in a closed form. To this end we use the following identities
\be
\sum _ {\ell=1} ^ \infty \f {e ^ {-t (\ell ^ 2 - 1/4)}} {(\ell-1/2) ^ s} = 2 ^ s \sum _ {k=1} ^ \infty \f {1} {k ^ s} e ^ {-t k (k + 2)/4} 
- \sum _ {k=1} ^ \infty \f {1} {k ^ s} e ^ {-t k (k + 1)} 
\ee
and
\be
\sum _ {\ell=1} ^ \infty \f {e ^ {-t (\ell ^ 2 - 1/4)}} {(\ell+1/2) ^ s} = 2 ^ s \sum _ {k=2} ^ \infty \f {1} {k ^ s} e ^ {-t k (k - 2)/4} 
- \sum _ {k=1} ^ \infty \f {1} {k ^ s} e ^ {-t k (k - 1)}.
\ee
Rewriting the sum
\be
\sum _ {\ell=1} ^ \infty \f {\ell ^ 2 \, e ^ {-t (\ell ^ 2 - 1/4)}} {(\ell+q) ^ s} = \sum _ {\ell=1} ^ \infty e ^ {-t (\ell ^ 2 - 1/4)} 
\l [ \f {1} {(\ell+q) ^ {s-2}} + \f {2(1-q)} {(\ell+q) ^ {s-1}} + \f {(1-q)^2} {(\ell+q) ^ {s}}\r ]  
\ee
and using the formula (\ref{heat_formula_1}) we obtain
\bea
\sum _ {\ell=1} ^ \infty {\ell ^ 2 \log \l ( \ell^ 2 - 1/2 \r )  e ^ {-t (\ell ^ 2 - 1/4)}} & = & - \f {\sqrt{\pi}} {16 \, t ^ {3/2}} \l ( \log 4 t + \g -2\r ) - \f {1} {8 \, t} - \f {\sqrt{\pi}} {64 \, t ^ {1/2}} \l ( \log 4 t + \g \r ) \nn \\ 
&& +\f {1} {96} \l (\g + \log t \r ) + \f {1} {12} \log 2 + \f {1} {4}\z ' (-1) - \f {9 \z (3)} {96 \pi ^ 2},
\label{F12}
\eea
and similarly
\bea
\sum _ {\ell=1} ^ \infty {\ell ^ 2 \log \l ( \ell^ 2 + 1/2 \r )  e ^ {-t (\ell ^ 2 - 1/4)}} & = & - \f {\sqrt{\pi}} {16 \, t ^ {3/2}} \l ( \log 4 t + \g -2\r ) + \f {1} {8 \, t} - \f {\sqrt{\pi}} {64 \, t ^ {1/2}} \l ( \log 4 t + \g \r ) \nn \\ 
&& +\f {1} {96} \l ( \g - \log t \r ) + \f {1} {24} \log 2 - \f {1} {4}\z ' (-1) - \f {9 \z (3)} {96 \pi ^ 2}.
\label{F32}
\eea
Putting all terms together produces the following result for the regularized partition function
\bea
F _ {S ^ 3} (t) =
- \f {\sqrt{\pi}} {8 \, t ^ {3/2}} \l ( \log 4 t + \g -2 \r ) - \f {\sqrt{\pi}} {32 \, t ^ {1/2}} \l ( \log 4 t + \g \r ) 
+ \f {1} {16} \l ( 2 \log 2 - \f {3 \z (3)} {\pi ^ 2} \r ).
\label{FS3_t}
\eea
It is clear that there are no covariant counterterms that can be used to remove divergent pieces in (\ref{F12}) and (\ref{F32}) separately, while for the sum (\ref{FS3_t}) the counterterms of the form (\ref{counterterms _S3}) will be sufficient, thus leading to the free energy (\ref{f_conf_scalar}).

\section{Conclusion \label{section_conclusion}}

It is not uncommon that naively computed expressions for certain quantities are divergent. However, predictions for observables should not depend on a regularization chosen to make sense of infinities. For instance, Ward identities (a manifestation of symmetries) should be satisfied. Some of regularizations preserve symmetries by construction others  demand symmetry breaking counterterms to restore the symmetry in the final result. The effective field theory perspective tells us in particular that whatever is not forbidden is possible, therefore, all admissible (consistent with symmetries and field/particle content) operators should be included in the Lagrangian. The advantage of regularizations respecting the symmetries in a manifest way is that in this case the space of possible operators in the Lagrangian is restricted, leading to the result in a more economic way.

In computing the Casimir energy or the partition function regularizations preserving diffeomorphism invariance are, for instance, zeta function or the dimensional regularizations. Another regularization (which is very much reminiscent of the zeta function one) was considered in the present paper. According to this regularization for an operator with a (discrete and positive) spectrum $\Lambda _ \a$ the finite value of the following sum
\be
\sum _ \a h(\Lambda _ \a),
\label{general_sum}
\ee
is given by the $t$-independent part of the sum regulated using a smoothed cut-off function~$\eta(x)$
\be
\mathrm {Finite} \, \sum _ \a h(\Lambda _ \a) \equiv \sum _ \a h(\Lambda _ \a) \eta (t \Lambda _ \a) \Big | _ {t-\text{indep}},
\ee
such that $\eta (0) = 1$ and it tends to zero sufficiently fast at infinity. This is well physically motivated, since divergent $1/t^{b}$ pieces can be associated with counterterms. The result does not depend on the choice of $\eta (x)$, but in the current paper we considered the case of $\eta (x) = e ^ {-x}$. This regularization in many cases allows to find the sum numerically in a straightforward and easy way. Let's consider eigenvalues labelled by only one quantum number $\lambda _ \ell$ with multiplicity $\m _ \ell$. Then the regulated sum becomes
\be
\sum _ \ell \m _ \ell \, h (\lambda _ \ell) e ^ {- t \lambda _ \ell}.
\ee
For large $\ell$ the summand can be expanded in a power\footnote{The terms with $\log \ell$ can be expressed as derivatives $\f {d} {ds} \ell ^{s}$.} series
\be
\m _ \ell f (\lambda _ \ell) \underset{\ell \to \infty}{=} \sum _ k g_ k \ell ^ {b_k} + O (\ell ^ {-2}).
\ee
Hence, the sum can be rewritten as follows
\be
\sum _ \ell  \sum _ k g_ k \ell ^ {b_k} e ^ {- t \lambda _ \ell} + \sum _ \ell \l [ \m _ \ell \, h (\lambda _ \ell) - \sum _ k g_ k \ell ^ {b_k} \r ],
\ee
where in the second term we dropped the factor $e ^ {-\lambda _ \ell t}$, since the sum is convergent and can be computed numerically. The asymptotic behavior of the first term can be found using the formulas (\ref{Riemann_Hurwitz_heat_formula}) or similar, thus, leading to the final result.

From this computation it is clear when an apparent misuse of the zeta function regularization, namely, a formal substitution
\be
\sum _ \ell  \sum _ k \ell ^ {b_k} = \sum _ k \z( -b_k ),
\label{zeta_formal}
\ee
may lead to a correct result (see examples in Section~\ref{section_zeta-function} and Appendix~\ref{Riemann_Hurwitz_heat}). If for large $\ell$ the eigenvalues have the following expansion
\be
\lambda _ \ell \underset{\ell \to \infty}{=} \ell ^ {a _ 0} + d _ 1 \ell ^ {a _ 1} + d _ 2 \ell ^ {a _ 2}+ \dots,
\ee
it is clear that the formal result (\ref{zeta_formal}) can be obtained if the sum (\ref{general_sum}) is regulated with $e^{-t \ell ^ {a_0}}$, rather than with (leading to the diffeomorphism invariance) $e ^ {-\lambda _ \ell t}$, since
\be
\sum _ \ell \ell ^ {b_k} \exp \l ( - t \ell ^ {a _ 0} \r ) = A t ^ {-(b _ k + 1)/ a _ 0} + \z( -b_k ).
\label{zeta_naive_non-diff}
\ee
The mismatch between the sums can be found as an asymptotic ($t \to 0$) expansion from the following integral
\be
t ^ {(b_k+1)/ a _ 0} \int dx \, x ^ {b_ k} \, \exp \l ( - x ^ {a_0} \r)  \l [ 1 +  \l (d_ 1 t ^ {1- a_1/a_0} x ^ {b_1} \r ) + 
\f {1} {2} \l (d_ 1 t ^ {1- a_1/a_0} x ^ {b_1} \r ) ^ 2 + \dots \r ]
\label{corrections_naive}
\ee
As a result the naive expression may be correct if none of the terms in (\ref{corrections_naive}) have $t ^ 0$ behavior thus not changing the 
$t$-independent part of the sum (\ref{zeta_naive_non-diff}).

\appendices

\section{Euler-Maclaurin formula  \label{Riemann_Hurwitz_heat}}

The Euler-Maclaurin formula is an extremely powerful tool approximating sums by integrals. Its derivation can be found in many textbooks, for instance~\cite{Hardy}. We use the following convention for the Euler-Maclaurin formula
\be
\sum _ {k=a+1} ^ b f(k) = \int_ a ^ b dx f (x) - B _ 1 \l [ f(b) - f (a) \r ] + 
\sum _ {k=1} ^ N \f {B_ {2k}} {(2k)!} \l [ f^ {(2k-1)} (b) - f^ {(2k-1)} (a) \r ] + R _ N,
\ee
where the remainder is given by
\be
R _ N = \int _ 0 ^ 1 \sum _ {k=a} ^ {b-1} f ^{(2N)} (t+k) B _ {2k}(t) dt,
\ee
with $f ^{k}(x)$ denoting the $k$-th derivative of the function $f(x)$ and $B_{k} (t)$ being Bernoulli polynomials defined as coefficients of the following Taylor series
\be
\f {x e ^ {xt}} {e ^ x-1} = \sum _ {k = 0} ^ \infty \f {B _ n (t)} {n!} x^n.
\ee
For $t=0$ the polynomials become Bernoulli numbers $B _ {k}(0) = B _ k$.

The formula (\ref{gen_theta_asympt}) allows to find the zeta function of the corresponding operator. For example for eigenvalues of the form 
$\lambda _ \ell ^ 0 = \ell$ or $\lambda _ \ell  ^ q = \ell+q$ choosing the cut-off function to be one of the two
\be
e ^ {- \lambda ^ {(0,q)} _ \ell t} ~~ \text{or} ~~ e ^ {- \sqrt {\lambda ^ {(0,q)} _ \ell} t},
\ee
it is straightforward to show using the Euler-Maclaurin formula that for $s<1$ the following asymptotic expansions hold (see for instance~\cite{TerryTao})
\bea
\sum _ {\ell=1}^\infty \ell ^ {-s} e ^ {-\ell t} & \underset{t \to 0} {=} & \f {\G (1-s)} {t ^ {1-s}} + \z (s) - t \, \z(s-1) + O (t^2), \\
\sum _ {\ell=1}^\infty \ell ^ {-s} e ^ {-\ell^2 t} & \underset{t \to 0} {=} & \f {\G \l (\f {1-s}{2} \r)} {2 \, t ^ {(1-s)/2}} + \z (s) - t \, \z (2-s) + O (t^2), \nn \\
\sum _ {\ell=0}^\infty \l ( \ell + q \r )^ {-s} e ^ {- \l ( \ell + q \r )t} & \underset{t \to 0} {=} & \f {\G (1-s)} {t ^ {1-s}} + \z (s,q) - t \, \z (s-1,q)+ O (t^2), \nn \\
\sum _ {\ell=0}^\infty \l ( \ell + q \r )^ {-s} e ^ {- \l ( \ell + q \r )^ 2t} & \underset{t \to 0} {=} & \f {\G (1-s)} {2 \, t ^ {1-s}} + \z (s,q)- t \, \z (2-s,q) 
+ O (t^2). \nn
\label{Riemann_Hurwitz_heat_formula}
\eea
In fact any term in the expansions above can be related to a derivative of the corresponding sum. For instance, for the coefficient in front of 
$t ^ {k+1}$, with $1 \leq k \in \mathrm Z$ one gets correspondingly
\be
\l [ \sum _ {\ell=1}^\infty \ell ^ {-s} e ^ {-\ell t} \r ] _ {k+1} = \f{(-1)^k}{k!}\z (k-s), ~~ \text{and} ~~ 
\l [ \sum _ {\ell=1}^\infty \ell ^ {-s} e ^ {-\ell^2 t} \r ] _ {k+1} = \f{(-1)^k}{k!} \z (2k-s).
\ee
Similarly one can prove that for $s<1$ the following asymptotic expansion is valid
\bea
\sum _ {\ell=1}^\infty \ell ^ {-s} e ^ {-\ell (\ell+q) t} \underset{t \to 0} {=} \displaystyle \f {t^{-\f {1-s} {2}}} {2} &\Bigg [& \G \l ( \f {1}{2}-\f{s} {2}\r ) 
- q \G \l ( 1- \f {s} {2} \r ) t ^ {1/2} + \f {q^2} {2} \G \l ( \f {3}{2}-\f{s} {2}\r ) t  \\ 
&& - \f {q^3} {3!} \G \l ( 2 -\f{s} {2} \r ) t ^ {3/2} + \f {q^4} {4!} \G \l ( \f {5}{2}-\f{s} {2}\r ) t^2 + 
O (t ^ 2) \, \, \, \Bigg ] + \z (s). \nn
\label{heat_formula_1}
\eea

Another example we want to show here is related to a misuse of the zeta function regularization. Imagine that the following sum should be computed for a system on a $2$-sphere~\cite{Hellerman:2015nra}
\be
\sum _ {\ell=1}^\infty (2 \ell + 1) \sqrt{\ell (\ell+1)}.
\ee
Formally it can be rewritten as
\be
\sum _ {\ell=1}^\infty \l ( 2 \ell^2 + 2 \ell + \f {1} {4} \r ) + \sum _ {\ell=1}^\infty \l [ (2 \ell + 1) \sqrt{\ell (\ell+1)} - 2 \ell^2 - 2 \ell - \f {1} {4}  \r ] 
= - 7/4 - 0.015. 
\label{conf_sphere_Q_naive}
\ee
At the same time regulating the sum according to the Section~\ref{section_conclusion} leads to
\bea
\sum _ {\ell=1}^\infty \l ( 2 \ell^2 + 2 \ell + \f {1} {4} \r ) e ^ {- t \ell(\ell+1)} +
\sum _ {\ell=1}^\infty \l [ (2 \ell + 1) \sqrt{\ell (\ell+1)} - \l ( 2 \ell^2 + 2 \ell + \f {1} {4} \r ) \r ] \nn \\
= \f {\sqrt{\pi}}{2 t ^ {3/2}} - 1/4 - 0.015. 
\label{conf_sphere_Q}
\eea
The mismatch is due to a nontrivial contribution to the $t$-independent part of the first term in (\ref{conf_sphere_Q}).

\section{Pauli-Villars regularization \label{Appendix_PV}}

In the main text we showed that the PV regularized partition function of a scalar coupled to a round $3$-sphere is given by (\ref{FS3}). 
One can show that using the Euler-Maclaurin formula the sum associated to the PV regulators in the limit $L \to \infty$ equals to
\bea
\sum _ {\ell = 1} ^ L \ell ^ 2  \log \l ([ \ell ^ 2 - 1 + \a _ {m} ^ 2(\x) a^2 + M^2_i a^2 \r ]
& = & \f{2L^3} {9} \l ( 3 \log L - 1 \r ) + L^2 \log L \\
&& + \l [ L a^ 2M _ i ^ 2 + L \l ( \f {L} {3} + \a _ m ^ 2(\x) a^2 + \f {5}{6}  \r )  \r ] \nn \\
&& + \l \{ -\f {\pi a^3 M _ i ^3} {3} + \f {a ^ 2 M_ i^2} {2} -\f {1} {2} \l [\pi a M_i - 1 \r ] \l [\a _ m ^ 2(\x) a^2 - 1 \r] \r \}. \nn
\eea
At the same time rewriting
\be
\log \l [ \ell ^ 2 - 1 + \a _ m ^ 2(\x) a^2 \r ] = \log \l ( \ell + i \sqrt{\a _ m ^ 2(\x) a^2 - 1} \r ) \l ( \ell - i \sqrt{\a _ m ^ 2(\x) a^2 - 1} \r ),
\ee
and using the formulas (\ref{asymptotic_Hurwitz_zeta}), (\ref{finite_two_zeta}) and the following relation
\be
\z ' (0,z) = \log \G (z) - \f {1} {2} \log 2 \pi, ~~ \mathrm {Re} z >0,
\ee
it is straightforward to show that in the limit $L \to \infty$
\bea
\sum _ {\ell = 1} ^ L \ell ^ 2 \log \l [ \ell ^ 2 - 1 + \a _ m ^ 2(\x) a^2 \r ] & = & \f {2 L^3} {9}  \l (3 \log L - 1 \r ) + L^2 \log L 
+ L \l ( \f {L} {3} + \a _ m ^ 2(\x) a^2 + \f {5}{6}  \r ) \nn \\
&& + F _ {S^3}(\sqrt{\a _ m ^ 2(\x) a^2 - 1}) - \f{1}{6} \l [ \a _ m ^ 2(\x) a^2 - 1\r ].
\eea
Combining the two results one gets the formula (\ref{PV_S3}) for the regularized free energy with
\bea
\label{finite_FS3_PV}
F _ {S^3} (x) = & - & \frac{1}{2} \l [  \zeta ^{'}(-2,1+i x) + \zeta ^{'}(-2,1-i x) \r ] \nn \\
&+& i x \l [ \zeta ^{'}(-1,1+i x) - \zeta ^{'}(-1,1-i x) \r ] \\
&+&\frac{1}{2} x^2 \l [ \zeta ^{'}(0,1+i x) + \zeta ^{'}(0,1-i x) \r ]. \nn
\eea
Although the regulator dependent pieces can be removed with the appropriate counterterms the function (\ref{finite_FS3_PV}) should not be considered as a prediction for the finite value of the partition function, for there is always an ambiguity in adding the counterterms of the form 
\be
b_1 m ^ 3 \mathrm {Vol} _ {S ^ 3} + b_2 m \mathrm {Ric} _ {S ^ 3}.
\ee

\subsection{General manifold $\mathbb M ^ 2 \times S ^ 1 _ \b$}

PV regularization can be used to find the free energy of a scalar field on an arbitrary manifold of the form $\mathbb M ^ 2 \times S ^ 1 _ \b$,
where $\b$ is the size of $S ^ 1$. Indeed, assuming that the eigenvalues of the corresponding operator on $\mathbb M ^ 2$ are 
$\Lambda _ \a > 0$ we get
\be
F = \f {1} {2} \sum _ {i=0} ^ N \sum _ {\a} \sum _ {k = - \infty} ^ \infty c _ i \log \l [ \l ( \f {2 \pi k} {\b} \r ) ^ 2 
+ \Lambda _ \a + M _ i ^ 2 \r ],
\ee
where $c_ 0 = 1$ is the non-regulator contribution ($M_0=0$) and $N$ is the number of regulators needed. Computing the sum in the way similar to the Section~\ref{PV_S1_ex} results in
\be
F = \f {\b} {2} \sum _ {i=1} ^ N \sum _ {\a} c _ i \sqrt {\Lambda _ \a + M _ \m ^ 2} + \sum _ \a  \log \l ( 1 - e ^ {- \b \sqrt {\Lambda _ \a} } \r ).
\ee
For instance in the case of a round $2$-sphere $S ^ 2 _ a \times S ^ 1 _ \b$ the procedure leads to
\be
F _ {S ^ 2 \times S ^ 1} = \sum _ {\ell = 0} ^ \infty (2 \ell + 1) \log \l \{ 1 - \exp \l [ - \f {\b} {a} \sqrt { (\ell + 1/2) ^ 2 + m ^ 2 a^ 2} \r ] \r \}.
\ee
while for a squashed $2$-sphere $S^2 _ {a,c} \times S^1 _ \b$ we obtain
\be
F _ {\text{sq}} = \sum _ {\ell = 0} ^ {\infty} \sum _ {k = -\ell} ^ {\ell} \log \l \{ 1 - \exp \l [ - \f {\b} {a} \sqrt { \l ( \ell + \f {1} {2} \r ) ^ 2 
+ k ^ 2 \l ( \f {a^2} {c^2} - 1 \r ) + m ^ 2 a ^ 2 }  \r ] \r \}.
\ee

\newpage
\bibliographystyle{utphys}
\bibliography{zeta_function}{}

\end{document}